\documentclass[12pt,manuscript,natbib209]{aastex}

\newcommand{\changed}{  }

\shorttitle{Survey for Giant Planets in Cold Debris Disks}
\shortauthors{Apai et al.}

\begin{document}
\newcommand{\ntargets}{8~}
\newcommand{\mjup}{M$_{\mathrm J}$~}

\title{A Survey for Massive Giant Planets in Debris Disks with Evacuated Inner Cavities\altaffilmark{1} }

\altaffiltext{1}{Based on observations collected at the European Southern Observatory at
Paranal, Chile (ESO programs P078.C-0412(A) and P077.C-0391(A))}

\author{D. Apai\altaffilmark{2}}
\affil{Steward Observatory, The University of Arizona, Tucson, AZ 85721, USA}
\altaffiltext{2}{Laplace Team, NASA Astrobiology Institute}
\email{apai@as.arizona.edu}

\and

\author{M. Janson}
\affil{Max Planck Institute for Astronomy, K\"onigstuhl 17, Heidelberg, D--69117, Germany}

\and 

\author{A. Moro--Mart{\' in}}
\affil{Department of Astrophysical Sciences, Peyton Hall, Ivy Lane, Princeton University, Princeton, NJ08544, USA} 

\and

\author{M.~R.~Meyer\altaffilmark{2}}
\affil{Steward Observatory, The University of Arizona, Tucson, AZ 85721, USA}

\and

\author{E. E. Mamajek}
\affil{Harvard--Smithsonian Center for Astrophysics, 60 Garden Street, 
MS-42, Cambridge, MA 02138, USA}

\and

\author{E. Masciadri}
\affil{INAF-- Astrophysical Observatory Arcetri, Italy}

\and

\author{Th.~Henning}
\affil{Max Planck Institute for Astronomy, K\"onigstuhl 17, Heidelberg, D--69117, Germany}

\and

\author{I. Pascucci, J. S. Kim}
\affil{Steward Observatory, The University of Arizona, Tucson, AZ 85721, USA}

\and

\author{L. A. Hillenbrand}
\affil{California Institute of Technology, Pasadena, CA 91125, USA}

\and

\author{M. Kasper}
\affil{European Southern Observatory, Karl--Schwarzschild--Str. 2, D--85748 Garching, Germany }

\author{B. Biller}
\affil{Steward Observatory, The University of Arizona, Tucson, AZ 85721, USA}


\begin{abstract}
The commonality of collisionally replenished debris around main sequence stars suggests that minor bodies
are frequent around Sun--like stars. Whether or not debris disks in general are accompanied by planets is yet unknown, but 
debris disks with large inner cavities -- perhaps dynamically cleared -- are considered to be prime candidates for hosting
large--separation massive giant planets.  
We present here a high--contrast VLT/NACO angular differential imaging survey for eight such cold debris disks. 
We investigated  the presence of  massive giant planets in the range of orbital radii  where the inner edge of the dust debris is expected. 
Our observations are sensitive to planets and brown dwarfs with masses  $>$3 to 7 Jupiter mass, depending
on the age and distance of the target star.  
 Our observations did not identify any planet candidates. {\changed We compare the derived planet mass upper limits to 
 the minimum planet mass required to dynamically clear the inner disks. While we cannot exclude 
 that single giant planets are responsible for clearing out the inner debris disks,  our observations constrain
 the parameter space available for such planets.}  The non--detection
 of massive planets in these evacuated debris disks further reinforces the notion 
 that the giant planet population is confined to the inner disk ($<$15~AU).
\end{abstract}



\section{Introduction}

Collisionally replenished debris dust surrounds about 10--20\% of the main sequence Sun--like stars (e.g. \citealt{Meyer2007}). Such widespread evidence for minor body collisions demonstrates that planetesimals orbit most stars. It is natural to ask whether or not rocky and giant planets  are also present in these systems. No convincing correlation could yet be found between close--in exoplanets and the presence of debris (e.g. \citealt{Amaya2007}, but  see \citealt{2005ApJ...622.1160B}). However, the presence of massive giant planets has been often invoked to account for the observed azimuthal or radial asymmetries at large radii in many debris disks (e.g. \citealt{Greaves2005,Wilner2002}).  While theory offers several alternative mechanisms (e.g. \citealt{Takeuchi2001,Wyatt2005}), dynamical clearing of dust parent bodies by giant planets remains a feasible and exciting theoretical possibility (e.g. \citealt{Amaya2005,Quillen2006,Levison2007, Morbidelli2007}).  

Examples for such possibly dynamically--cleared disks include two recently identified disks 
around the young Sun--like stars \object{HD 105}  \citep{Meyer2004}  and \object{HD 107146} \citep{Williams2004}. 
Both disks were found to exhibit strong excess emission at wavelengths longer than 30~$\mu$m, while
displaying no measurable excesses shortward of 20~$\mu$m. The detailed analysis of the spectral energy distribution of  \object{HD 105} suggests that it is consistent with a narrow dust ring ($<$4~AU) with an inner radius of $\sim$42~AU, if the dust grains emit like black bodies \citep{Meyer2004}. 
Using a similar model \citet{Williams2004} showed that the excess emission from \object{HD 107146} is consistent with arising from cold dust (T=51~K) emitting as a single--temperature black body. The lack of measurable infrared excess shortward of 25~\micron ~illustrates that the inner disk regions are well cleared of dust: for HD~107146 there is at most 140$\times$ less warm dust (T=100~K) than cold  dust (T=51~K, \citealt{Williams2004}). The findings of the spectral energy distribution model for \object{HD 107146} have been confirmed
by direct imaging with the Hubble Space Telescope, that strengthen the case for a large featureless dust ring outside of
an evacuated inner cavity \citep{Ardila2004}.

Recent high--contrast imaging surveys have hinted on the general scarcity of giant planets at such large separations 
(e.g. \citealt{Masciadri2005,Kasper2007,Biller2007,Lafreniere2007}). 
Quantitative statistical analysis of the non--detections demonstrates that -- at a 90\% confidence level -- 
the giant planet population cannot extend beyond 30~AU if it follows a $r^{0.2}$ radial distribution, consistent
with the radial velocity surveys. The statistical analysis suggests an outer cut--off  for the giant planet 
population at $<$15~AU \citep{Kasper2007}. If so, dynamically cleared cold debris disks
may be the ssignposts for rare large--separation giant planets, ideally suited for direct imaging studies.

In this paper we report on a VLT/NACO high--contrast imaging survey for large--separation giant planets 
around  HD~105, HD~107146, and six other similar disks. In the following we will review the target stars 
and disks, the observations, followed by a comparison of our non--detections to lower planet mass 
limits set by dynamical clearing simulations.

\subsection{Targets}
Our targets were selected from the sample of 328 Sun--like stars (0.7--2.2 $M_\odot$) targeted in  the {\em Formation and Evolution of Planetary Systems}  Spitzer Space Telescope Legacy program (FEPS, \citealt{Meyer2006}). 
From this sample we identified \ntargets southern stars, which: a) display strong infrared excess emission at 
long wavelengths ($\lambda > 20 \mu$m); b) no measurable excess emission at shorter wavelengths; and, 
C) are young  and close enough to permit the detection of planetary--mass objects within the inner radius of the cold debris. 
Table~\ref{T:Targets} gives an overview of the key parameters of the target stars. The typical lower mass limit for
the debris in the systems is $10^{-4}$ to $10^{-5}$~M$_{\earth}$, making these disks massive analogs of our 
Kuiper--belt (\citealt{Meyer2004,Kim2005}, Hillenbrand et al., in prep). {\changed The disks of \object{HD 105} and  \object{HD 107146} ---
included in our sample --- have  inner evacuated regions with an estimated radii of $\sim40$~AU and 
$\sim31$~AU  \citep{Meyer2004,Williams2004}. The other six disks exhibit spectral energy distributions similar to
 \object{HD 105} and \object{HD 107146}. Based on the similarity of the excess emissions and the almost identical
spectral types all eight disks are expected to have cleared--out inner disks of similar size.}
The only possible exception in this sample is \object{HD 202917}, for which the re--calibration of the IRAC fluxes
after our VLT observations revealed a faint, but likely real infrared excess even at wavelengths shortward of 10\micron,
suggesting that this inner disk may harbor small, but non--negligible amounts of warm dust.

In the following we discuss briefly the results of the age determination for these sources as this has direct impact on the sensitivity
of our observations to giant planets. A more detailed discussion of the ages of the whole FEPS sample 
will be presented in Hillenbrand et al. (in prep). We briefly summarize the upper and lower age estimates ($t_{min}$ and $t_{max}$) for each star along  with the most likely age $t_{prob}$, where available.
 HD~105 has already reached the main sequence ($t_{min}$=27~Myr) and its chromospheric activity suggests a $t_{max}$ of 225~Myr (Hillenbrand et al., in prep.). Very likely a member of the Tuc--Hor moving group \citep{Mamajek2004} its $t_{prob}$ is 30~Myr \citep{Hollenbach2005}. 
HD~377 is also a main sequence star ($t_{min}>25$~Myr) and the chromospheric activity suggests that $t_{max}$=220~Myr. 
The median of four other age indicators sets $t_{prob}$=90~Myr.
For HD~107146  we adopt  the age range of 80--200~Myr. HD~202917 is a 
likely member of the Tuc--Hor moving group ($t_{min}=t_{prob}=30$~Myr) and its upper age limit is set by its Li--abundance, higher than that of the Pleiades ($t_{max}<100$~Myr). HD~35850 is suggested to be a $\beta$~Pic Moving Group member ($t_{min}$=12~Myr, \citealt{Song2003}) and its observed rotation rate sets a reliable upper age limit of $t_{max}$=100~Myr (Hillenbrand et al. in prep.; cf. 
\citealt{Barnes2007}). 
HD~70573 is among the few stars that are known to harbor both a debris disk and a giant planet. \citet{Setiawan2007} found an $m_2 sin i = 6.1$~\mjup possible planet on a 1.76--AU orbit. A combination of different age indicators suggest a $t_{min}=30$~Myr for HD~70573 and a $t_{prob}=60~$Myr; \citet{Setiawan2007} quotes $t_{max}$=125~Myr. 
{\changed Based on Li--abundance and chromospheric activity, position on the color--magnitude diagram and the analysis of its space motions Mamajek et al. (in prep.)  estimates that HD~209253 has $t_{min}=200$~Myr and $t_{max}<1.6$~Gyr  with $t_{prob}$=500~Myr.}
HD 25457 is a member of the AB~Dor moving group giving a very strong lower age limit ($t_{min}=50$~Myr, \citealt{Zuckermanetal2004}). \citet{Luhman2005} derives an age of 75--125~Myr (we adopt $t_{prob}=75$~Myr), while the upper age limit is set by the chromospheric activity ($t_{max}=170$~Myr).

\begin{deluxetable}{lccccccc}
\tabletypesize{\scriptsize}
\tablecaption{Target parameters. \label{T:Targets}}
\tablewidth{0pt}
\tablehead{\colhead{Target} &  \colhead{R.\,A. (J2000)} & \colhead{Dec. (J2000)}  &  \colhead{V--mag.$^a$}& \colhead{Dist. [pc]$^a$} &\colhead{Sp. Type} & \colhead{Ages:$^b$ $t_{min}$/$t_{prob}$/$t_{max}$}  }
\startdata
HD 105    & 00 05 52.6 & $-$41 45 11 & 7.51 & 40 & G0V    & 27 Myr / 30 Myr / 225 Myr\\
HD 377    & 00 08 25.7 & $+$06 37 01 & 7.59 & 40 & G2V    & 25~Myr / 90~Myr / 220~Myr  \\
HD 25457  & 04 02 36.8 & $-$00 16 08 & 5.38 & 19 & F5V    & 50 Myr / 75 Myr / 170 Myr\\
HD 35850  & 05 27 04.8 & $-$11 54 03 & 6.30 & 27 & F7/8V  &  12 Myr /  12 Myr / 100 Myr\\
HD 70573  & 08 22 50.0 & $+$01 51 34 & 8.69 & 70 & G1/2V  & 30 Myr / 60 Myr / 125 Myr \\
HD 107146 & 12 19 06.5 & $+$16 32 54 & 7.04 & 29 & G2V    &  80 Myr /  -- / 200 Myr \\
HD 202917 & 21 20 50.0 & $-$53 02 03 & 8.65 & 46 & G5V    &   30 Myr / 30 Myr / 100 Myr \\
HD 209253 & 22 02 33.0 & $-$32 08 02 & 6.63 & 30 & F6/7V  &  200~Myr / 500 Myr / 1.6 Gyr \\
\enddata
\tablenotetext{a}{All magnitudes and distances from the Hipparcos catalog, except for the distance
of HD~70573, which is a main sequence--distance. }
\tablenotetext{b}{The age estimates are discussed in the text.}
\end{deluxetable}

\section{Observations and Data Reduction}

Our \ntargets targets were observed with ESO's Very Large Telescope using the NACO adaptive optics system \citep{Lenzen2003,Rousset2003}.
The observations were carried out in service mode in late 2006 and early 2007. The weather conditions were
excellent with typical visual seeing of 0\farcs8 and clear skies.

We used the spectral differential imaging mode (SDI) of NACO in order to enhance the contrast for any methane--rich cold {\changed (T$<1200$~K)} companion (e.g. \citealt{Lenzen2004}). The SDI mode uses two Wollaston prisms to split the incoming light rays into four beams of nearly identical light path. These rays pass through four narrow--band filters, two of which are identical. The three different filters ($f_1$, $f_2$ and $f_3$ corresponding to 1.575, 1.600, and 1.625~$\mu$m) probe the 1.62~$\mu$m  methane feature and the adjacent continuum. Because the SDI mode uses the 1024$\times$1024--pixel S13 camera of NACO with a $13" \times 13"$ field of view, the simultaneous acquisition of 4 images in this field reduces the effective field of view to about $3" \times 3"$. 

{\changed The achieved contrast, however, is not as good as predicted, probably due to the combined effect of read--out noise and a slightly lower Strehl ratio. With the reduced contrast our observations were sensitive only to planets  beyond the 1--3 \mjup planet mass range. These
planets -- at the young ages of out targets -- are too hot to display the 1.62 $\mu$m--methane feature  \citep{Burrows2003}. The lack of the methane feature results
in almost identical planet fluxes in the $f_1$, $f_2$ and $f_3$ filters, rendering the SDI technique inefficient.}

Instead, we opted to reduce the data taken in the $f_1$ filter ($\lambda_c=1.575 \mu$m, $\delta\lambda=0.025 \mu$m) in the angular differential imaging mode, i.e. without applying the spectral differential imaging step  (e.g.,  \citealt{Mueller1987,Marois2006,Kasper2007}). 
The data reduction was performed with a dedicated  pipeline, as described in detail in \citet{Kellner05} and \citet{Janson07}.  The frames taken at a given rotator angle were averaged and the collapsed frame corresponding to one angle was subtracted from the other. 
This procedure cancels out residual static or quasi--static features from the instrument, whereas any companion will 
remain as a combination of a positive and negative point source. 
The intensity of the residuals (at a certain  separation from the primary) is characterized by taking the standard deviation in a $9 \times 9$--pixel  square {\changed (0\farcs11$\times$0\farcs11 $ \approx (2.7 \lambda/D)^2$ )} centered on that separation, at 180 evenly sampled angles, and taking the median of the results. 
This is repeated for all separations to create a radial profile of the error distribution in the image. When combined with the  brightness of the primary, this yields the achieved contrast as a function of separation in the final image (see, \citealt{Janson07} for more  details). {\changed Previous artificial planet tests on identical data sets processed with the reduction pipeline used here showed that  3--sigma--bright sources would have been reliably identified as {candidate} planets \citep{Janson07}. Lacking any such detection, we used 3--sigma fluxes as
upper limits on the brightness of any companions to the target stars.}
In order to convert the achieved contrast in $f_1$ to the more commonly used H--band, we derived a conversion factor by  comparing the flux densities in the two filters in a simulated spectrum of a giant planet in the age and mass range probed by our  observations \citep{Burrows2003}.

\begin{deluxetable}{lcccccc}
\tabletypesize{\scriptsize}
\tablecaption{Log of the observations.  \label{T:Observations}}
\tablewidth{0pt}
\tablehead{\colhead{Target} & \colhead{UT Dates}   &   \colhead{NDIT $\times$ DIT$^a$}  & \colhead{Frames per Angle} &  \colhead{Angles} & \colhead{On--Source} &   \colhead{Strehl} }
\startdata
HD 105 	   & 07/20/06+08/18/06+08/21/06 &    1 $\times$ 4s & 144 & 0\degr, 33\degr & 19 min & 43\%\\
HD 377    &  08/27/06  &   1 $\times$ 4s & 144 & 0\degr, 33\degr & 19 min & 37\%\\
HD 25457 & 08/12/06+08/13/06 & 1 $\times$ 4s& 144 & 0\degr, 33\degr &19 min & 46\%\\
HD 35850   & 08/13/06 &  1 $\times$ 4s& 144 & 0\degr, 33\degr & 19 min& 40\%\\
HD 70573   & 03/02/07 &   21 $\times$ 5s & 16 & 0\degr, 33\degr & 56 min & 58\%\\
HD  107146    & 04/260/06+0 05/26/06 & 1 $\times$ 4s & 95 & 0\degr, 33\degr &13 min & 51\%\\
HD 202917 &  06/23/06 & 1 $\times$ 12s & 32 & 0\degr, 33\degr &13 min & 45\%\\
HD 209253    &  07/09/06+07/16/06+07/23/06 & 1 $\times$ 4s & 144 & 0\degr, 33\degr &19 min & 49\%\\
\enddata
\tablenotetext{a}{DIT -- Detector integration time; NDIT -- number of integration averaged on--chip. }
\end{deluxetable}

\section{Results}

The NACO/ADI observations acquired high--contrast, high--resolution images of the
 \ntargets{} target stars and their immediate environment (typically between 10 and 70~AU). 
In spite of the  sensitive observations we could not identify giant planet candidates or any 
other point sources in the images. 

\begin{figure}
\epsscale{1.0}
\plotone{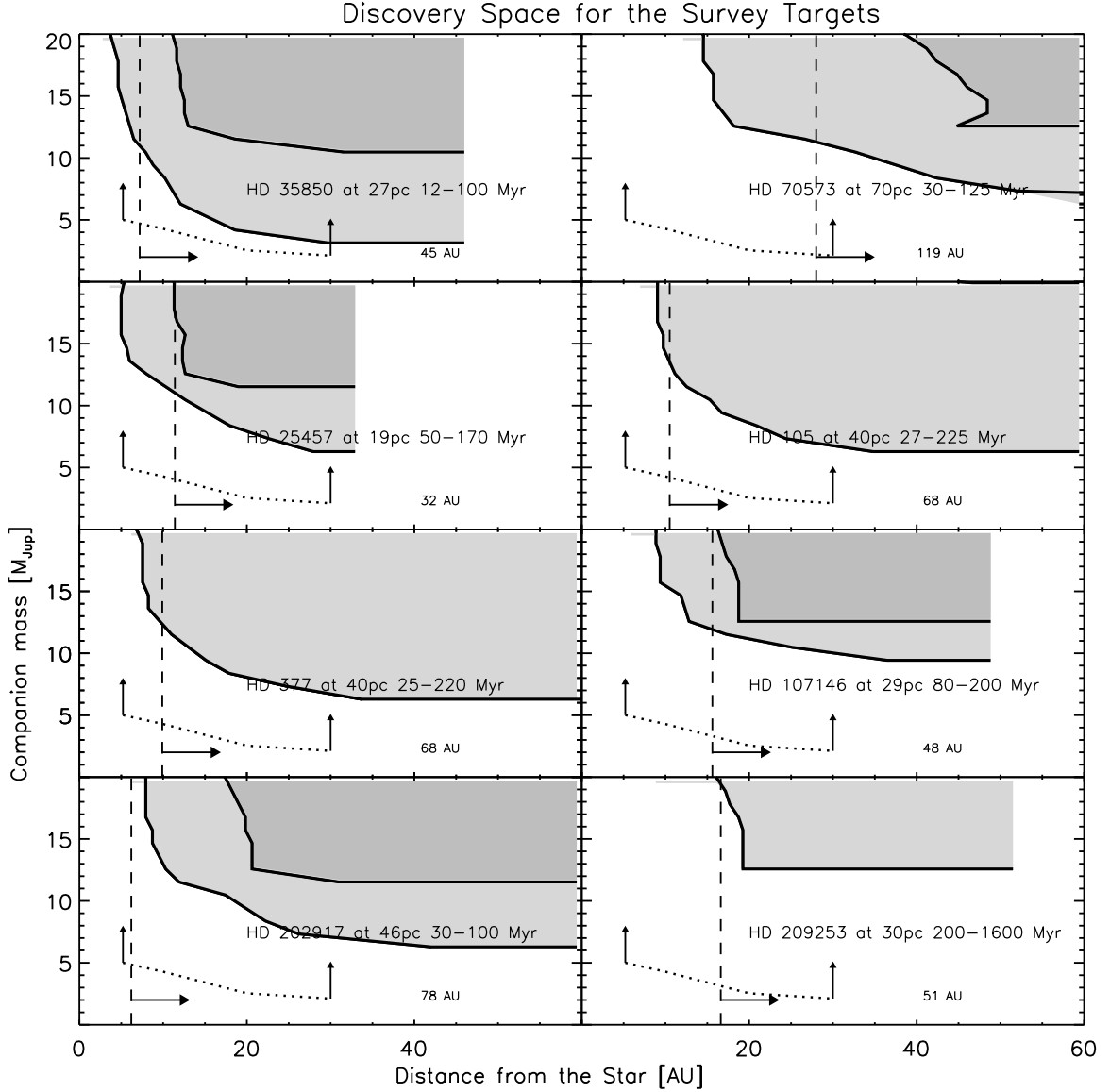}
\caption{The discovery space of the NACO observations (shaded), the minimum predicted planet masses required
for a single planet to scatter out $>90\%$ of the planetesimals at a given radius (dotted lines) and the lower limits
for the disk inner radius (dashed lines, Hillenbrand et al., in prep.). The presence of planets within the shaded parameter range is excluded by our observations. 
The upper and lower shaded sensitivity curves mark the limits for a possible younger and older stellar age. 
For each figure the outer radius of the field of view is given in astronomical units in the lower right--hand corner.
Note, that the reversing sensitivity curve for HD~70573 and HD 209253 is due to the reversal of the planet mass--luminosity curve in the corresponding evolutionary phase \citep{Baraffe2003}.
 \label{Sensitivities}}
\end{figure}

Our data analysis allows us to set firm upper limits to the brightness of the sources that would have 
been identified as a candidate. We use the age estimates in Table~\ref{T:Targets} and planetary evolution models
\citep{Baraffe2003} to convert the achieved sensitivities to planet masses.
These limits are shown in Fig.~\ref{Sensitivities} as a function of separation from the target stars for
the lower and upper age limits of our stars. 

Using the high--contrast observations we can study the probable range of radii in these disks at which the dust debris 
and the parent body planetesimals reside. Our measurements exclude the presence of any brown dwarf companions 
for virtually all our targets at these radii. For HD~209253, the oldest star in our sample,  the images exclude companions
down to the brown dwarf/giant planet mass boundary (13~\mjup) at radii 20~AU or greater.
For the the youngest source HD 35850 (12--100~Myr) 
our observations exclude any giant planet companions down to 3--4 \mjup between 25 to 45~AU, if the star 
belongs to the $\beta$~Pic moving group as suggested by \citet{Song2003}.
For the other six sources our observations are typically sensitive to $\sim$6~\mjup at orbital radii $>$30--40~AU. 

{\changed Note, that the dominant uncertainty of the upper limits stems from the difficulty of stellar age determination and from
the poorly constrained initial conditions for giant planet evolution models. In particular, if shocks 
lead to efficient energy dissipation during the accretion phase, giant planets may start with much lower  
luminosities (e.g. \citealt{Marley2007}) than assumed by the hot--start models 
(e.g. \citealt{Baraffe2003,Burrows2003}).}

\section{Discussion: Inside--Evacuated Debris Disks Without Massive Giant Planets?}

If single planets are responsible for clearing out the inner disks in the observed systems, they will be located
very close to the inner edge of the debris. Although the available data does not allow the direct measurement
of the inner disk radii, simple black body fits to the spectral energy distributions  provide
reliable {\em lower} limits. We adopt these limits from Hillenbrand et al. (in prep.) and note that they range
from 6.2 to 28~AU. Fig.~\ref{Sensitivities} shows the limits for the individual sources (dashed lines).
We stress again that these are lower limits --- the real inner disk radii are probably somewhat larger.

Given the range of detectable planet masses in Fig.~\ref{Sensitivities}, we assess whether
dynamical clearing by a less massive and therefore undetectable planet could
still be a feasible mechanism to explain the lack of measurable quantities 
of warm dust, or on the contrary, if the dynamical clearing scenario may be 
rejected. Because our disks are devoid of gas \citep{Pascucci2006}  we use the dynamical models by \citet{Amaya2005}
to investigate the effect of a giant planet on the dust population. These models
investigate the efficiency of dust particle ejection by gravitational scattering as a 
function of planet mass and planet location. In these models the dust particles 
are released from an outer belt of planetesimals and drift inward toward the central 
star under the effect of Poynting--Robertson drag, scattering as they cross 
the orbit of the planet and naturally creating a dust--depleted region inside its orbit. 
However, as shown in the upcoming study by Hillenbrand et al. 
(in prep.) all our disks are collision--dominated, as are bright debris disks 
in general \citep{Wyatt2005}. This means that the dust particles may not have
time to drift too far from the parent bodies before getting eroded by collisions
down to the blow-out size. Thus, in these collision--dominated disks the dust generally
traces the location of the planetesimals. Therefore, we need to evaluate the effect of 
a giant planet on the planetesimal population rather than on the dust particles. 
Because gravitational scattering is a process independent
of mass, the models of \citet{Amaya2005} are also applicable to planetesimals 
as long as these can be considered to  be ``test particles'' (i.e. their masses are negligible with respect to that of the planet). 

Using the above models we evaluate what is the mass of the least massive planet
that can open a gap in the planetesimal distribution. In order to provide
a good model for the evacuated inner gaps we require that the 
planet scatters out at least 90\% of the planetesimals. A dotted line in Fig.~\ref{Sensitivities} shows
these lower masses as a function of orbital radius.
We find that for planets in the 5--30 AU range, a planet mass of {\em at least} 2--5 \mjup masses 
is required. Thus, it is conceivable that giant planets in the mass range 2 to 5~\mjup clear the
gaps and still remain undetected by our survey. However, we point out that in the cases of
most of our targets, and in particular for  HD~35850, the parameter space
that such a planet can occupy is {\changed limited}.

{\changed Given that our high--contrast imaging survey did not find single, large--separation giant planets that may be responsible for
clearing the inner disks, we briefly explore alternative mechanisms. }
 \citet{Besla2007} provides a useful summary of the proposed models and
we will only highlight here a particularly interesting proposal  proposed recently by \citet{Amaya07} for the system HD~38529,  where two close--in planets trigger secular resonances that affect the planetesimal population in 
the outer disk. In this model the eccentricity of the planetesimals at the location of the secular
resonances is excited, thus enhancing the rate of collisions and truncating the planetesimal disk.



Given the range of possible mechanisms that may lead to the formation of dust rings it
is probable that in--depth studies of the individual systems will be required for judging
the feasibility of the proposed models on a case--by--case basis. {\changed However, our non--detections show 
that the presence of evacuated large inner holes in cold disks could be related to localized dust production or 
concentration of dust grains by dust--gas interactions rather than dynamical clearing by single massive giant planets.}


\section{Conclusions}

We present results from a high--contrast angular differential imaging survey of  \ntargets 
cold debris disks, selected to have significantly or totally evacuated inner disks.
Our observations searched for massive giant planets that may be  responsible for 
carving out the inner holes in the  observed cold debris disks. 
For most of our targets we reach typical sensitivities of  3 to 7 \mjup 
between 20 to 50~AU separations, but did not identify any likely planet
or brown dwarf candidates.

{\changed By comparing the derived planet mass upper limits to lower limits derived from 
dynamical scattering models (typically 2-5 \mjup between 10 and 30 AU), 
we limit the parameter space available for any single planet capable of 
efficiently clearing out the inner planetesimal disks. }

Our survey complements recent direct imaging surveys of nearby young stars indicating 
that massive giant planets at large separations are very rare. Cool debris disks
with large inner evacuated cavities remained promising possible exceptions to this
rule until now. However, the combination of  our observational upper limits and 
theoretical lower limits strongly suggest that massive giant planets at large separations 
are not present in most of these systems, reinforcing the finding that the outer 
cut--off for the giant planet distribution is probably at 15~AU or at even smaller 
semi--major axes \citep{Kasper2007}.

\acknowledgments

We thank the staff at the Paranal Observatories for the support of the service mode 
observations. In particular, we are grateful to S. Mengel and G. Lowell--Tacconi for their help with
the preparation of the observations. This material is partly based upon work supported by the 
National Aeronautics and Space Administration through the NASA Astrobiology Institute under Cooperative 
Agreement No. CAN-02-OSS-02 issued through the Office of Space Science to the Life and Planets 
Astrobiology Center (LAPLACE). We would like to thank members of the FEPS team for their help in 
characterizing the target stars and their disks. FEPS is pleased to acknowledge support through NASA 
contracts 1224768, 1224634, and 1224566 administered through JPL. 

{\it Facilities:} \facility{VLT (NACO)}.


\bibliographystyle{aa}
\bibstyle{aa}

\bibliography{lit}

\end{document}